\documentclass[preprint,12pt]{elsarticle}

\usepackage{amssymb}
\usepackage{tikz}
\usepackage{units}
\usepackage{natbib}
\usepackage{lineno}

\journal{Nuclear Instruments and Methods in Physics}

\begin{document}
\begin{frontmatter}

\title{Measuring the Attenuation Length of Water in the CHIPS-M Water Cherenkov Detector}

 
\author{F.Amat$^1$, P.Bizouard$^1$, J.Bryant$^2$, T.J.Carroll$^{5}$, S. De Rijck$^5$, S.Germani$^2$, T.Joyce$^3$, B.Kriesten$^4$, M.Marshak$^3$, J.Meier$^3$, J.K.Nelson$^4$, A.J.Perch$^{2}$, M.M.Pf{\"u}tzner$^{2}$, R.Salazar$^5$, J.Thomas$^2,^6$, J.Trokan-Tenorio$^6$, P.Vahle$^4$, R.Wade$^7$, C.Wendt$^6$, L.H.Whitehead$^2$, M.Whitney$^3$}

\address{$^1$Aix Marseille University Saint-Jerome, 13013 Marseille, France}
\address{$^2$Dept. of Physics and Astronomy, UCL, Gower St, London WC1E 6BT,UK}
\address{$^3$School of Physics and Astronomy, University of Minnesota, Minneapolis, MN 55455, USA}
\address{$^4$Department of Physics, College of William \& Mary, Williamsburg, Virginia 23187, USA}
\address{$^5$Department of Physics, University of Texas at Austin, Austin, Texas 78712, USA}
\address{$^6$Department of Physics, University of Wisconsin, Madison, WI 53706, USA}
\address{$^7$Avenir Consulting, Abingdon, Oxfordshire, UK}

%\ead{jennifer.thomas@ucl.ac.uk}

\begin{abstract}

The water at the proposed site of the CHIPS water Cherenkov detector has been studied to measure its attenuation length for Cherenkov light as a function of filtering time.  
A scaled model of the CHIPS detector filled with water from the Wentworth 2W pit, proposed site of the CHIPS deployment, in conjunction with a 3.2\,m vertical column  filled with this water, was used to study the transmission of  405\,nm laser light.
Results consistent with attenuation lengths of up to 100\,m were observed for this wavelength with filtration and UV sterilization alone.
\end{abstract}


\end{frontmatter}

%\begin{linenumbers}

\section{Introduction}
The CHIPS (CHerenkov Detectors In PitS) R\&D project will investigate the proof of principle for a novel and very inexpensive water Cherenkov detector \cite{ref:fnalloi}.
Demonstration of a cost in the region of \$200-\$300\,k per kilo-ton (kt) is one of the goals of this program of research in order to realize mega-ton scale detectors in the future within realistic financial constraints.
A prototype detector of between 3-5\,kt of water is proposed for construction 7\,milli-radians (mrad) off-axis in the NuMI neutrino beam, in a flooded mine pit 708\,km from the neutrino source at Fermilab. The prototype detector will observe neutrinos via the detection of Cherenkov radiation produced by their interactions in the water inside the detector. 

A major part of the cost saving in the CHIPS concept is achieved by enclosing the detector water volume in a very lightweight, water- and light-tight containment vessel to be submerged in the Wentworth pit. The 60\,m depth provides an overburden of water to provide cosmic ray shielding at the same time as physically supporting the detector volume. This avoids the need to build a large structure or to excavate many kilotons of rock to contain the target water. The Cherenkov light from neutrino interaction remnants in the water will be collected by planes of 80\,mm photomultiplier tubes (PMTs) positioned on the inner downstream side of the detector. It is therefore essential that the attenuation length of the Cherenkov light in the water be long enough to ensure very little reduction from production point to PMT. The pit once functioned as an open taconite mine and it is expected that the water will contain substantial suspended particulates. These can be removed by conventional filtering. The goal of this study was to understand whether, after filtering the remaining dissolved solids in the water would provide significant attenuation to the light. 
If this were to be case, further treatment of the water such as reverse osmosis or de-ionisation would be necessary, potentially leading to significant density differences between the inner and outer water, as the purified pit water will be used as the inner detector target mass. This could, in turn, lead to a significant increase in the cost, not only of the water treatment system, but also of the outer structure itself. The purification process and the method of the attenuation length measurement  are reported here.

The attenuation length of the water from the Wentworth pit, made using a Secchi disk \cite{ref:secchi}, is in the region of 2.5$\pm$0.5\,m.   
In order to achieve the 50-100\,m attenuation length required for a very large diameter detector \cite{ref:leighandandy}, the proposal is to fill the detector with pit water that has been passed through a series of filters down to a Micron Rating of 0.2\,$\rm{\mu m}$ .

\section{Experimental Setup}

Pit water contained in a 50 (U.S.) gallon (189 litre) barrel is continuously passed through 
the cleaning system comprised of  a UV sterilizer, a 1\,$\mu$m carbon filter, and a 0.2\,$\mu$m general purpose filter at a rate of 0.06 litres per minute (lpm) and circulated back into the barrel. 
The rate is set to simulate the rate of turnover in a large 30\,m diameter CHIPS detector which will circulate 490\,lpm for each 2\,m high detector slice.
Generally, about 7 turnovers are required in order to ensure that 99\% of the water has actually passed through the filters~\cite{ref:waterturnover}. This  would lead to a full water exchange approximately every two weeks.

%stuff from richard
%so for a 2 day turnover with a 2m slice of a 30m diameter detector i calculated that we need a suction hose of at least 3.5inches
%So here is the theory. whenever you use a pump to pull water up from a reservoir (or pit in our case) the hardest you can suck is about one atmosphere. We have to overcome two %pressure drops in the hose. The fist is the lift from the pit surface to the pump and the second is the drop in pressure along the 1000ft of the hose caused by friction and turbulence. %The lift from the pit is a few meters (maybe 3 say). You then calcuate the pressure drop de to the flow and for 129gpm through a 3.5in pipe this works out to be about half and %atmosphere.

\begin{figure}
  \centering
    \includegraphics[width=0.44\textwidth]{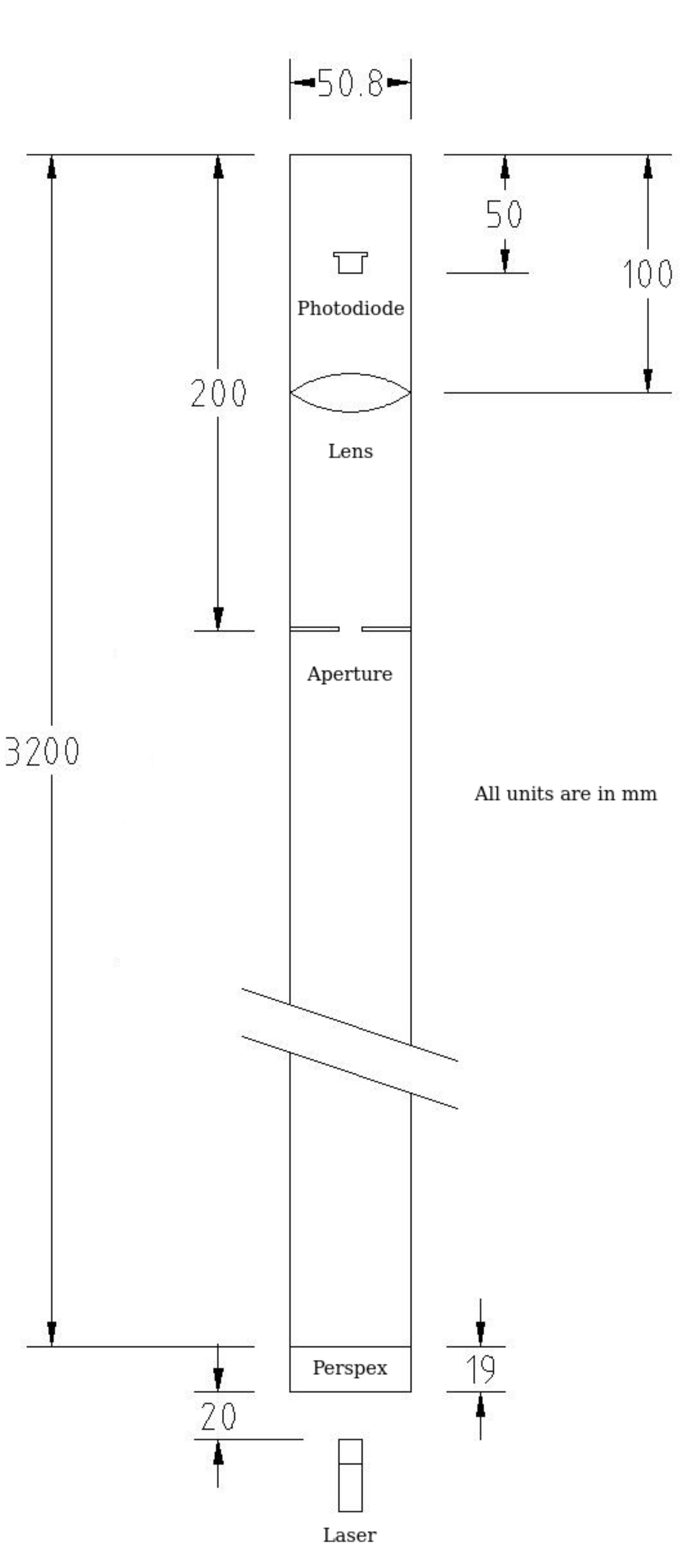}
    \includegraphics[width=0.40\textwidth]{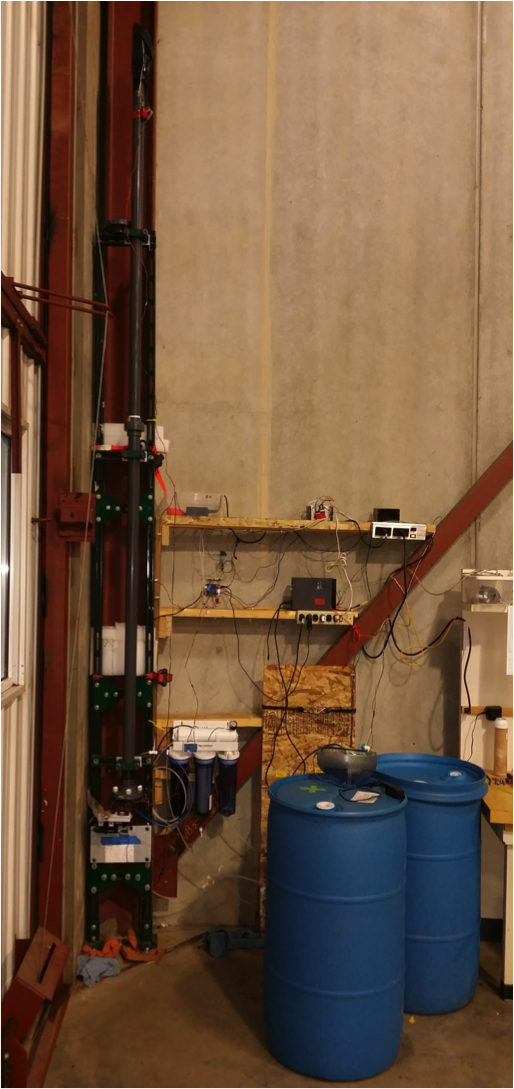}
        \caption{Left: schematic showing the key features of the water column (lengths in mm) and right: photograph of the setup in the laboratory showing the column (left), the filters(centre) and the barrel (forefront right).} \label{fig:layout}
        \end{figure}

A vertically mounted cylindrical tube of 3.2\,m length and 50.8mm (2~\,inch) inner diameter schedule 80 PVC pipe was sealed with a transparent perspex plate at the bottom through which a single 6\,mm water pipe was inserted. This setup was inspired by the measuring equipment at University of California, Irvine \cite{ref:michealsmy} and is shown in Figure \ref{fig:layout}.

%
%\begin{figure}[t]
%\begin{minipage}[t]{0.6\textwidth}
%\begin{center}
%\vspace{-7.5cm}
%\begin{tikzpicture}
%   \node[anchor=south west,inner sep=0] (image) at (0,0,0) {\includegraphics[width=3in]{watercolumn}};
 %   \begin{scope}[x={(image.south east)},y={(image.north west)}]
   %       \draw (0.75,0.8) -- +(0.3in,-0.2in)node[anchor=west] {lens};
      %    \draw(0.67,0.545) -- +(0.2in,-0.1)node[anchor=west] {Water Level};
         % \draw (0.45,0.95) -- +(-0.5in,-0.1)node[anchor=east] {PIN diode};
          %\draw (0.45,0.1) -- +(-0.5in,-0.1in)node[anchor=east] {Laser};
   % \end{scope}
%\end{tikzpicture}
%\end{center}
%\end{minipage}
%\begin{minipage}{0.4\textwidth}

%\vspace{-2.2cm}
%\includegraphics[width=2.5in]{watercolumnphoto}
%\end{minipage}

%\end{figure}

\begin{figure}
  \centering
    \includegraphics[width=0.68\textwidth]{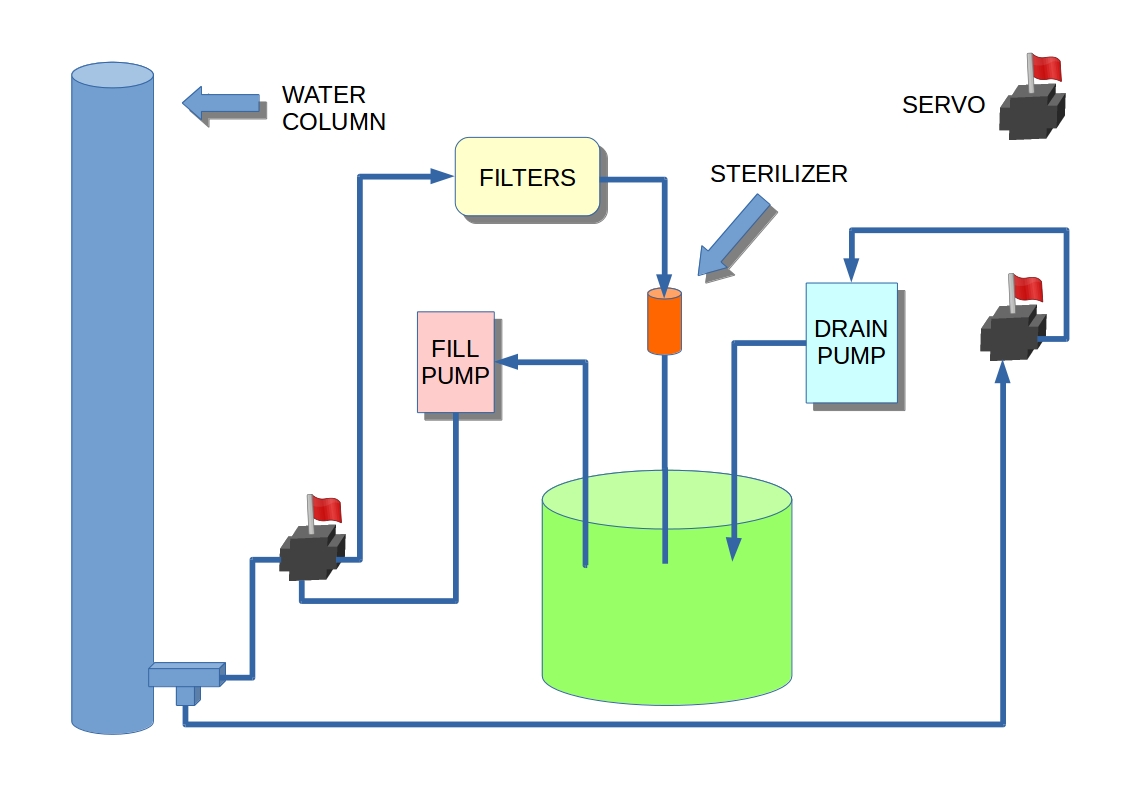}
  \caption{Sketch showing the layout of the pipework for the water circulation and the attenuation length measurement. The water continuously circulates until a measurement is performed at which point the servo direction is changed such that the column is filled before turning off the fill pump.}
\label{fig:pipework}
\end{figure}

The fully automated measuring apparatus was controlled by a BeagleBone Black (BBB) \cite{ref:BBB} single-board computer running Linux.  
The column was filled with the water from the barrel via a servo operated valve controlled by one of the BBB Pulse Width Modulation pins to direct the water away from the filters and into the tube.  The schematic of the water system is shown in Figure~\ref{fig:pipework}.
The fill pump was allowed to run until the column contained a total water height of 3m, and was then stopped via a relay controlled by one of the BBB GPIO pins.
A Thorlabs CPS405 collimated laser diode was mounted on a stage which allowed for adjustments in rotational and translational planes and was positioned to shine vertically upwards along the axis of the column. A lens with a focal length of 50\,mm was inserted at the top of the column, 50\,mm below a blue sensitive photo diode (1125-1009-ND 525NM) on axis at the top of the tube looking down at the laser.

\begin{figure}
  \centering
    \includegraphics[width=1.0\textwidth]{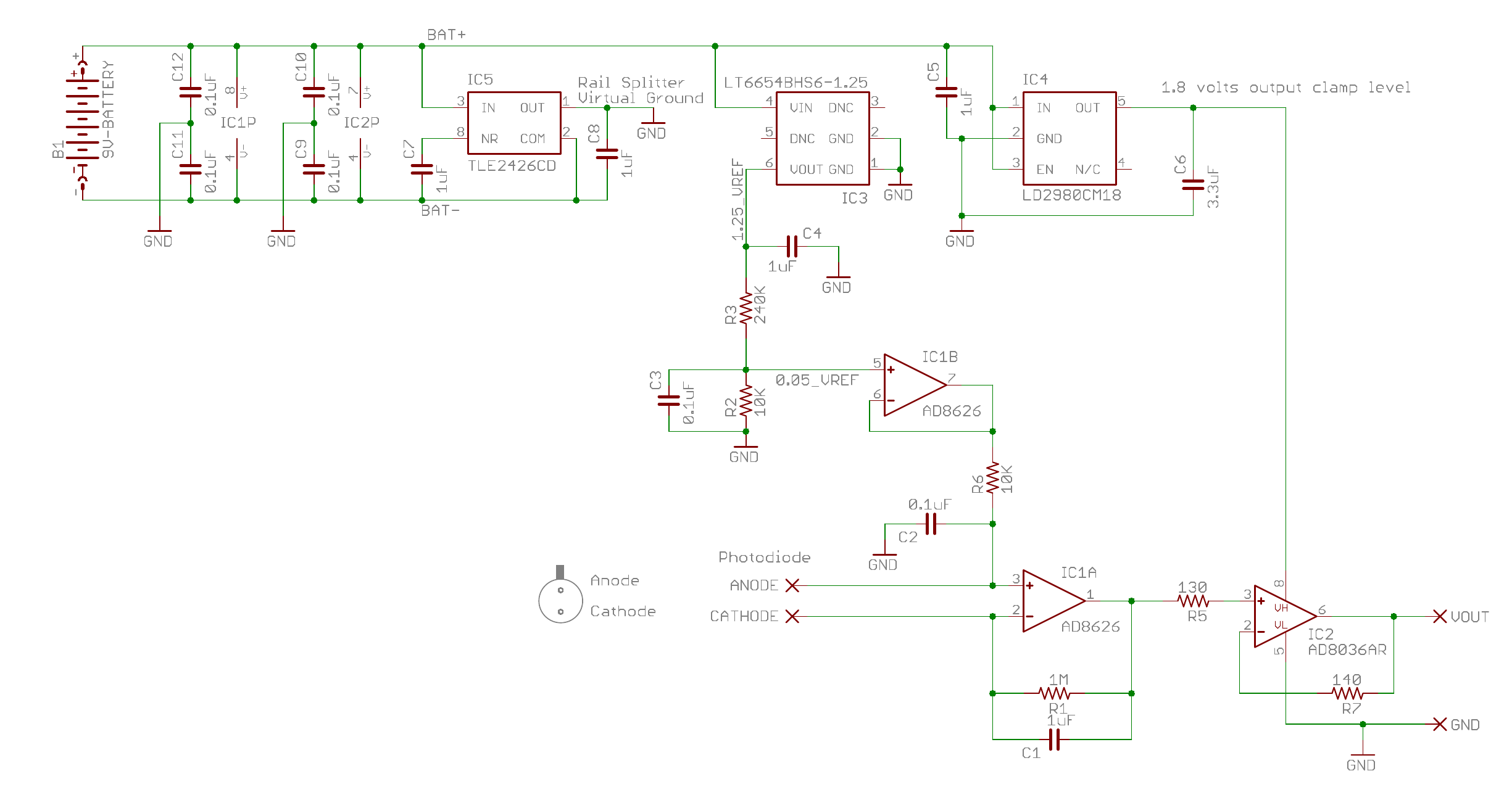}
    \includegraphics[width=0.40\textwidth,angle=90]{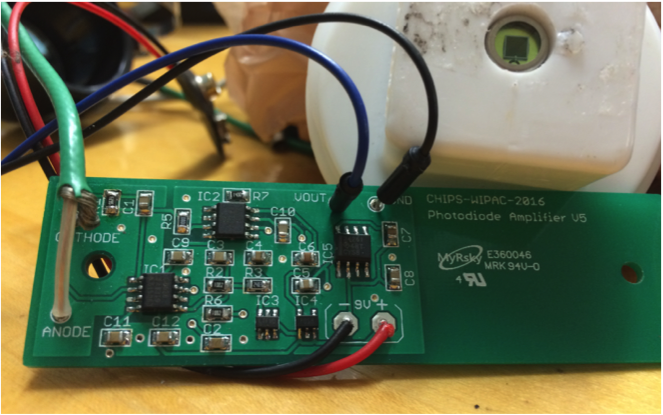}
  \caption{Amplifier circuit diagram (top) and fabricated board + photodiode (bottom)}
  \label{fig:pincircuitdiag}
\end{figure}

The photodiode output was amplified using an operational amplifier circuit as shown in Figure \ref{fig:pincircuitdiag}.  It was connected to the inputs of a trans-impedence amplifier, with feedback resistors of between 1 and 10\,M$\Omega$ depending on the required gain so that the photodiode was operated in short circuit mode and not reverse biased. There was a voltage offset such that the output rested at around 50\,mV when the input current was zero. A clamping output buffer amplifier prevented the output from going above 1.8\,V to avoid damage to the BBB input. The power source was a 9\,V battery.
 
\begin{figure}
  \centering
    \includegraphics[width=0.45\textwidth]{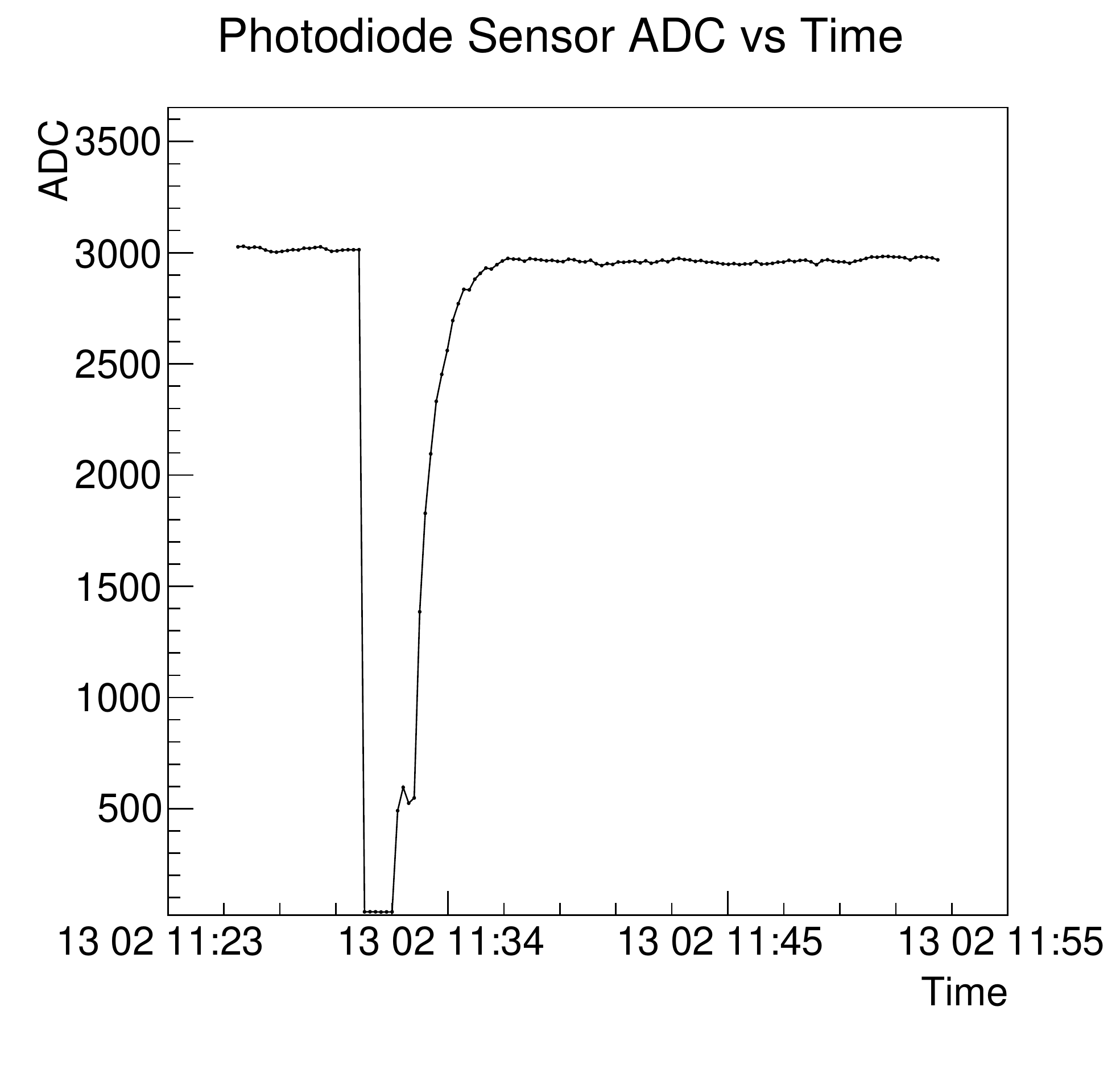}
    \includegraphics[width=0.45\textwidth]{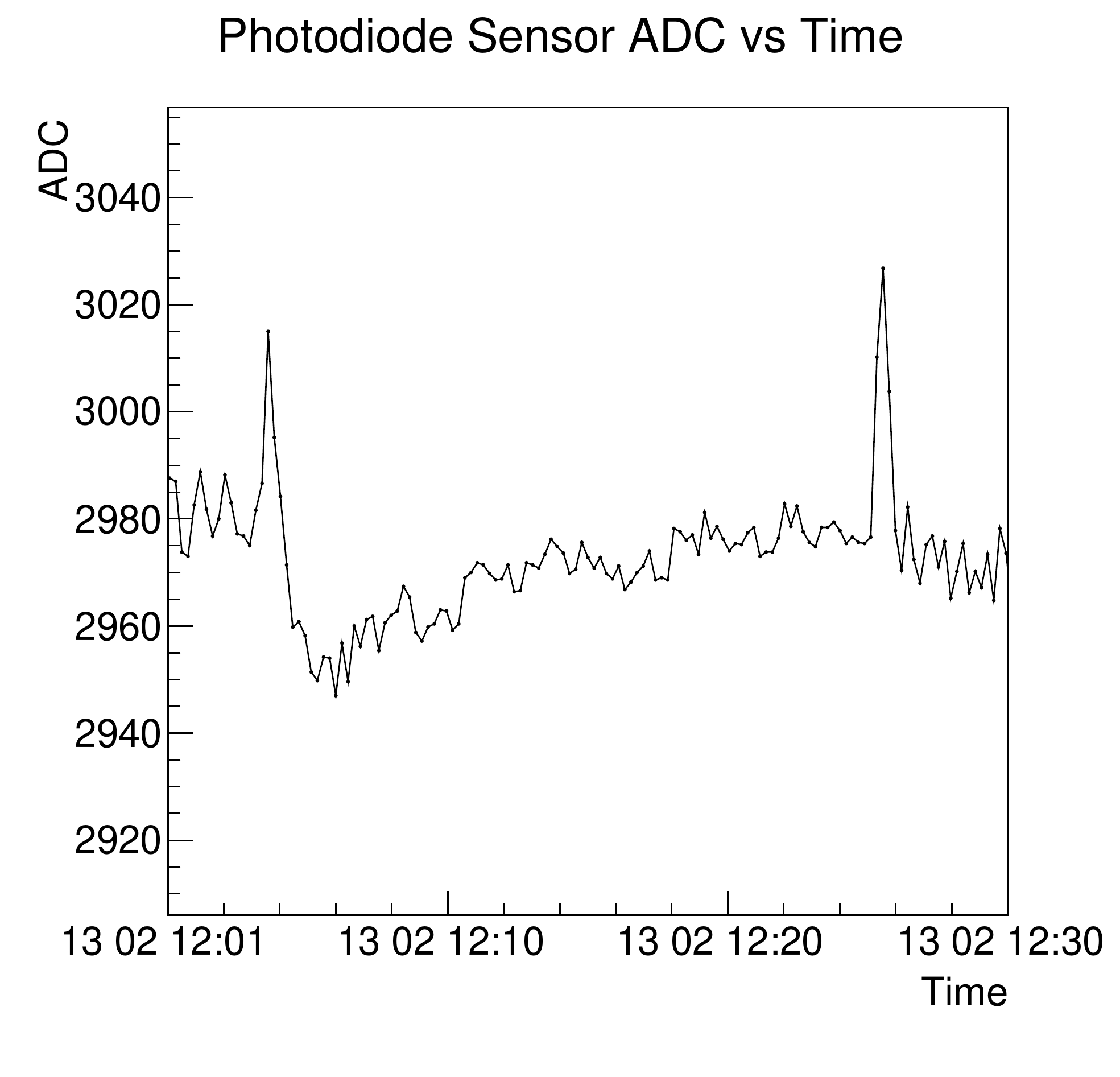}
  \caption{The left plot shows the photodiode ADC as a function of time. The fill is occurring where the ADC value has fallen to zero and the turbulence renders the water completely opaque. The right plot shows a time range after the fill is complete.  When  the drain pump turns on and off, the pipe shakes and this vibration causes turbulence in the water for a short time as evidenced by the peaks. }
  \label{fig:recovertime} 
\end{figure}

Once the column was full of water, the laser light was monitored in order to estimate the time it would take for the bubbles to rise and the water 
to become stable after the turbulence caused by the fill pump. Figure~\ref{fig:recovertime} (left) shows the photodiode ADC as a function of time. The fill 
can be seen to occur where the ADC value falls to zero. A conservative period of 20 minutes was taken to allow for recovery. This is the time period from the 
fill point to the end of the x axis in the Figure. The light level
can be seen to stabilize after this time. The next step in the process is to pump out a given amount of water. This also adds instantaneous vibration to the system, as shown in
Figure \ref{fig:recovertime} (right) where the two peaks correspond to the drain pump turning on and off and the recovery of the water clarity is seen in between.
The BBB DAQ  script was programmed to wait for 20 minutes after filling and at each measurement point. 

The output from the photodiode circuit was digitized with one of the ADC channels on the BBB and data recorded in ASCII files for analysis. A total of 50 measurements was taken at each water height over a period of about 10 seconds. A quantity of water was then pumped out of the column and the procedure of pausing and taking data was repeated.
The attenuation length was determined by measuring the intensity of light from the laser incident on the photodiode as a function of the height of water in the column and then performing an exponential fit to the graph. The error on the water height steps was less than 1\,cm; the fill time and emptying time were very repeatable. The number of steps could be varied by changing the amount of time the drain pump ran between data taking steps.

The water was circulated continuously through the filtration system  in between taking regular attenuation length measurement runs.
Improvements in the attenuation length were tracked as a function of time and therefore the effectiveness of the filtration and UV systems could be investigated. 

\section{Data Analysis}

The intensity of the light at the top sensor is  given by
\begin{equation}
\rm{I=I_{0}e^{-x/\lambda}}
\end{equation}
where I$_{0}$ is the light intensity with no water in the column, $\lambda$ is the attenuation length and x is the water height. The data are fit
to extract the attenuation length.

The accuracy of the measurement is limited by the length of the column, the stability of the laser and the photodiode, the stability of the ambient temperature, the vibrational stability of the environment, the stability of the BBB ADC and the amplifier circuit noise. There is therefore a limit to the maximum attenuation length measurement achievable. 

The attenuation length in the 100\,kt detector is required to be at least 30\,m at 405nm, as determined by reconstruction studies done on simulated neutrino interactions \cite{ref:leighandandy} but  a longer attenuation length translates into more light at the PMTs and therefore better reconstruction of the interaction. Clearly, the longer the column, the more sensitive the measurement can be as it will increase the difference between the light levels when empty and full.

\begin{figure}
  \centering
    \includegraphics[width=0.6\textwidth]{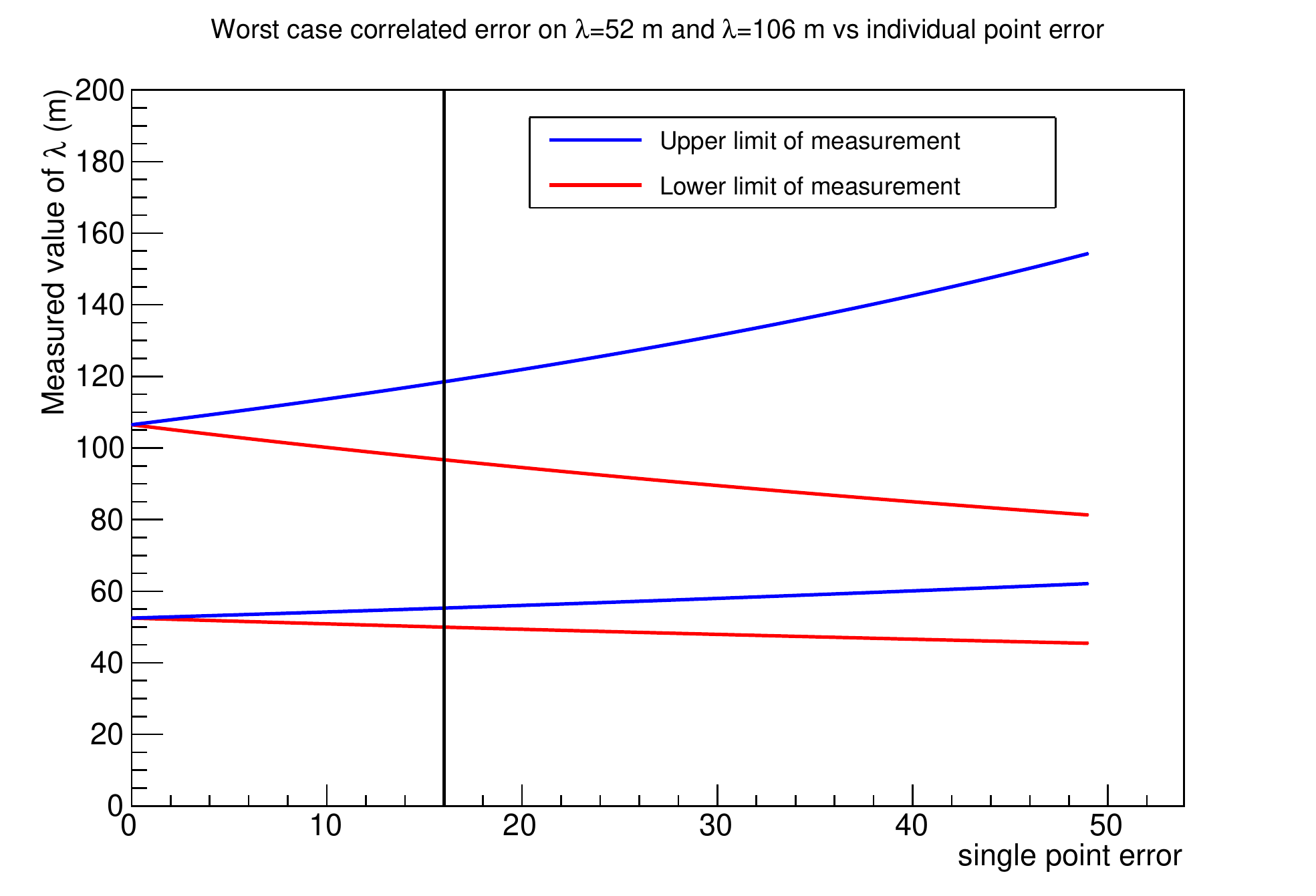}
  \caption{An illustration of the effect of the single point uncertainty on the measurement of the attenuation length, assuming a true attenuation length of 52\,m and 106\,m. In the case of a 16 ADC count error on each point (as measured in Section \ref{sec:improve}), shown by the vertical line, the attenuation length could be mis-measured by up to 2.5\,m and up to 12\,m respectively}
  \label{fig:illustrate} 
\end{figure}

As an illustration, two examples are investigated.
Taking the empty light level to be 3600 ADC counts in order to use most of the dynamic range of the 12 bit ADC, an attenuation length of 106\,m would correspond to a drop in intensity to 3500 ADC counts (3\%) over a 3\,m long length of water when the column was full. Similarly, an attenuation length of 52\,m would correspond to a drop in intensity of 6\% and the light level would measure 3400 ADC counts when full. Each attenuation length measurement involves N$\approx$10 separate measurements of the light level between full and empty which leads to an expectation of the uncertainty on the gradient  to be the error on a single point divided by $\sqrt{N}$. With very long attenuation lengths ($>$50\,m) a linear approximation works for the purposes of this illustration. Figure \ref{fig:illustrate} shows the possible range of uncertainty produced by the pipe length limitation on the true attenuation length of 52\,m and 106\,m as a function of the single point error. The red lines show the lower limit and the blue lines show the upper limit. This provides a sanity check that a measurement of 100\,m is safely far from the limitation of the 3\,m column if the combined systematic error per point is kept at below about 30 ADC counts. Each contribution to the systematic error was measured as described below.

\subsection{Systematic Errors}
\begin{figure}
  \centering
    \includegraphics[width=0.7\textwidth]{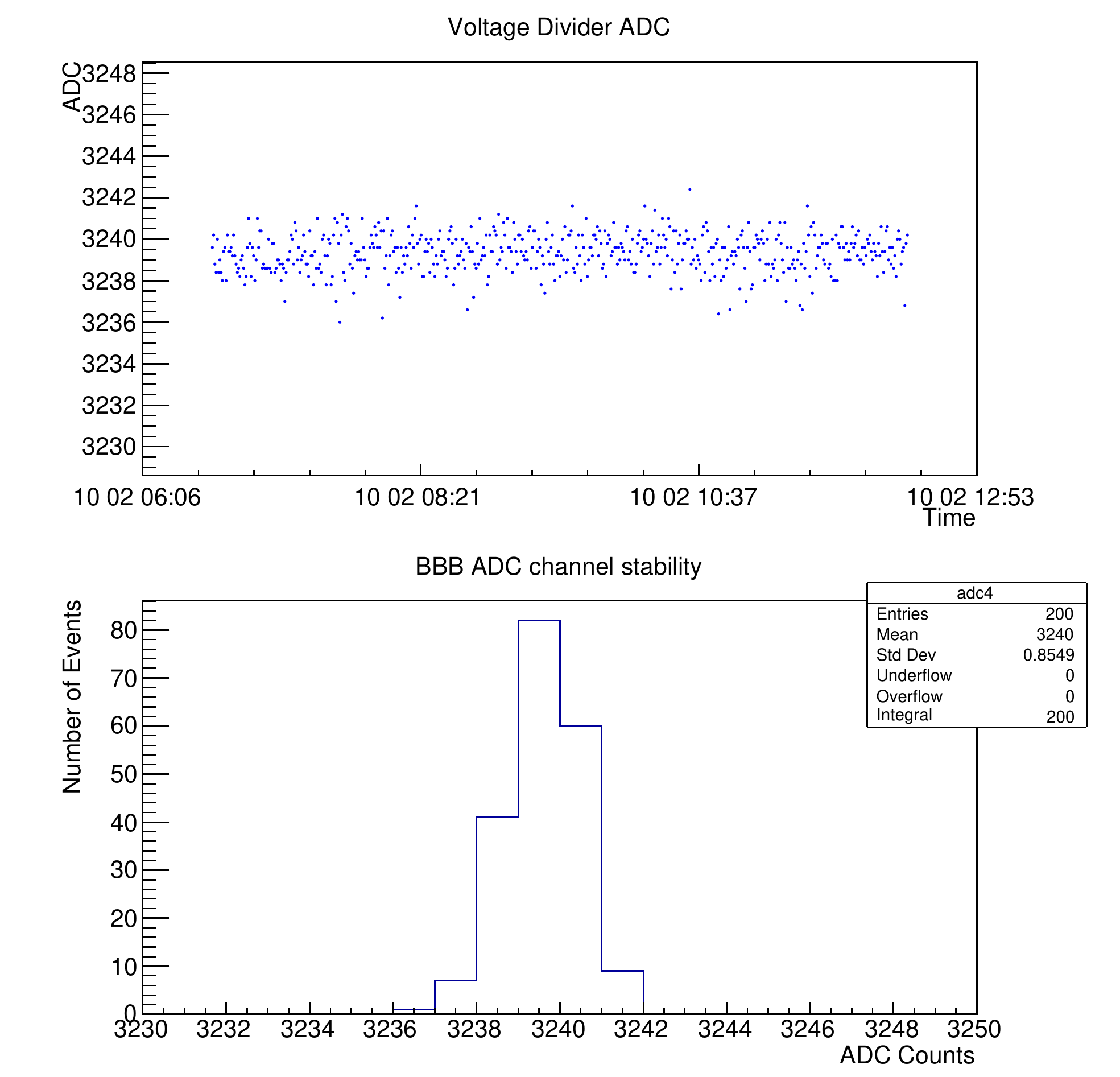}
  \caption{Top: BBB ADC channel reading from  a voltage divider as a function of time. Bottom: the projection of the time evolution showing a standard deviation of 1.1 ADC counts.}
  \label{fig:BBBstability}
\end{figure}

The contributions from the various sources of systematics were measured.
As shown in Figure~\ref{fig:BBBstability} the BBB ADC was measured to give a purely random jitter with a standard deviation of 1.1 ADC counts  by measuring one of the BBB ADC channels connected to a simple voltage divider circuit.  
The laser specification gives a stability of the laser output power of 1\% over a 1 minute period and  2\% over an 8 hour period. The attenuation length
measurement run time is between one and two hours, so a systematic of 1.5\% for the overall stability during this time period might be expected.

\begin{figure}[h]
  \centering
    \includegraphics[width=0.6\textwidth]{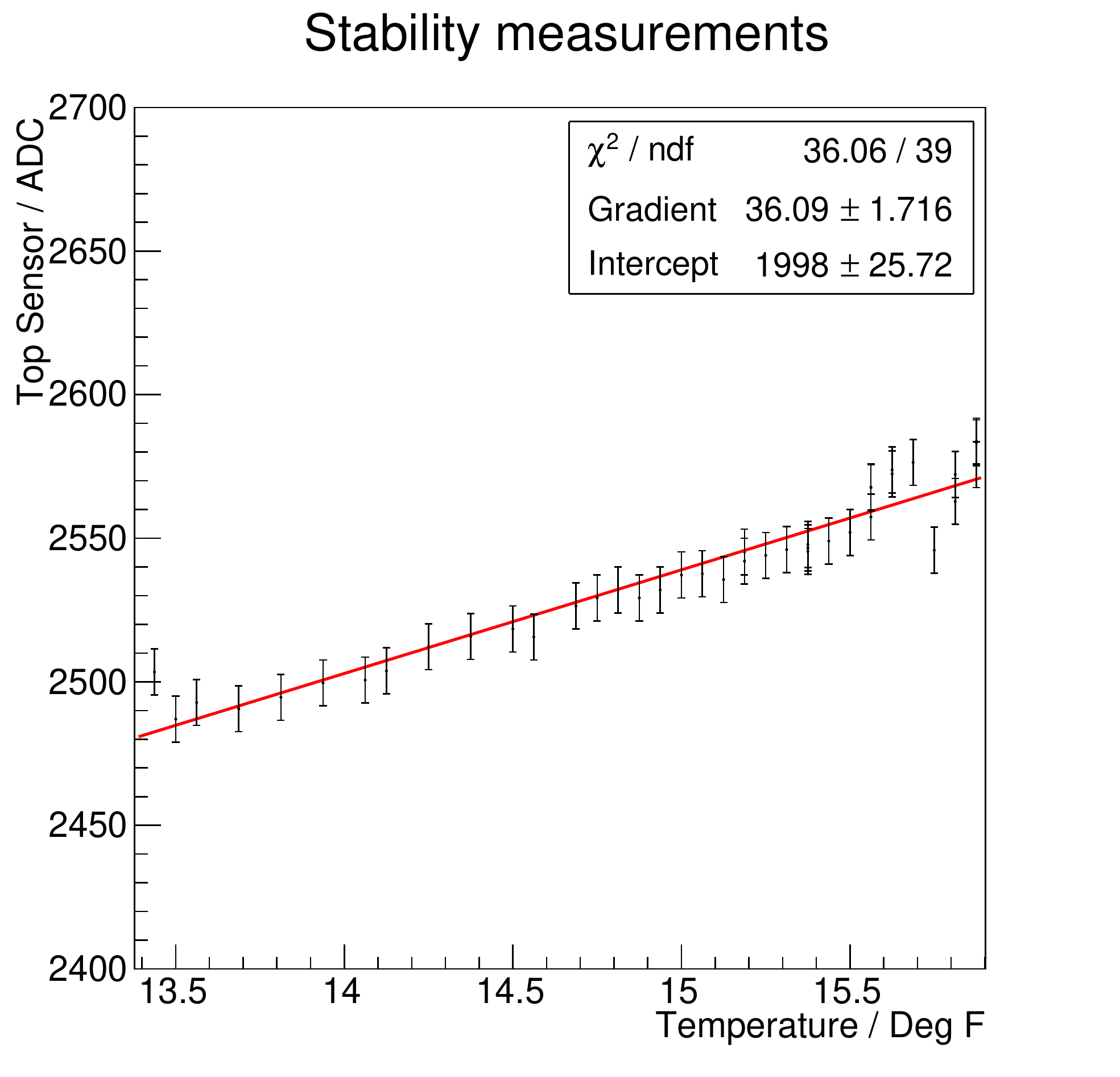}
  \caption{Top: Photodiode ADC as a function of temperature  
  for a full water column taken over a period of a few hours. A linear fit shows a change of about 38 ADC counts per $^{o}$C. Continuous temperature monitoring was used to estimate the size of the  errors on the data points for a given attenuation length measurement run.}
  \label{fig:tempvadc}
\end{figure}

The photodiode and laser stability are measured together by monitoring the ADC output as a function of time for a full water column. In addition ambient vibrations
which affect the full water column  are folded in to this measurement.
Figure~\ref{fig:tempvadc} shows the dependence of the photodiode output with temperature over a period of a few hours. 
The temperature was found to have the largest impact on the system stability of all the systematics. 

\subsection{Cleaning Rate Measurement}
\begin{figure}
  \centering
    \includegraphics[width=0.45\textwidth]{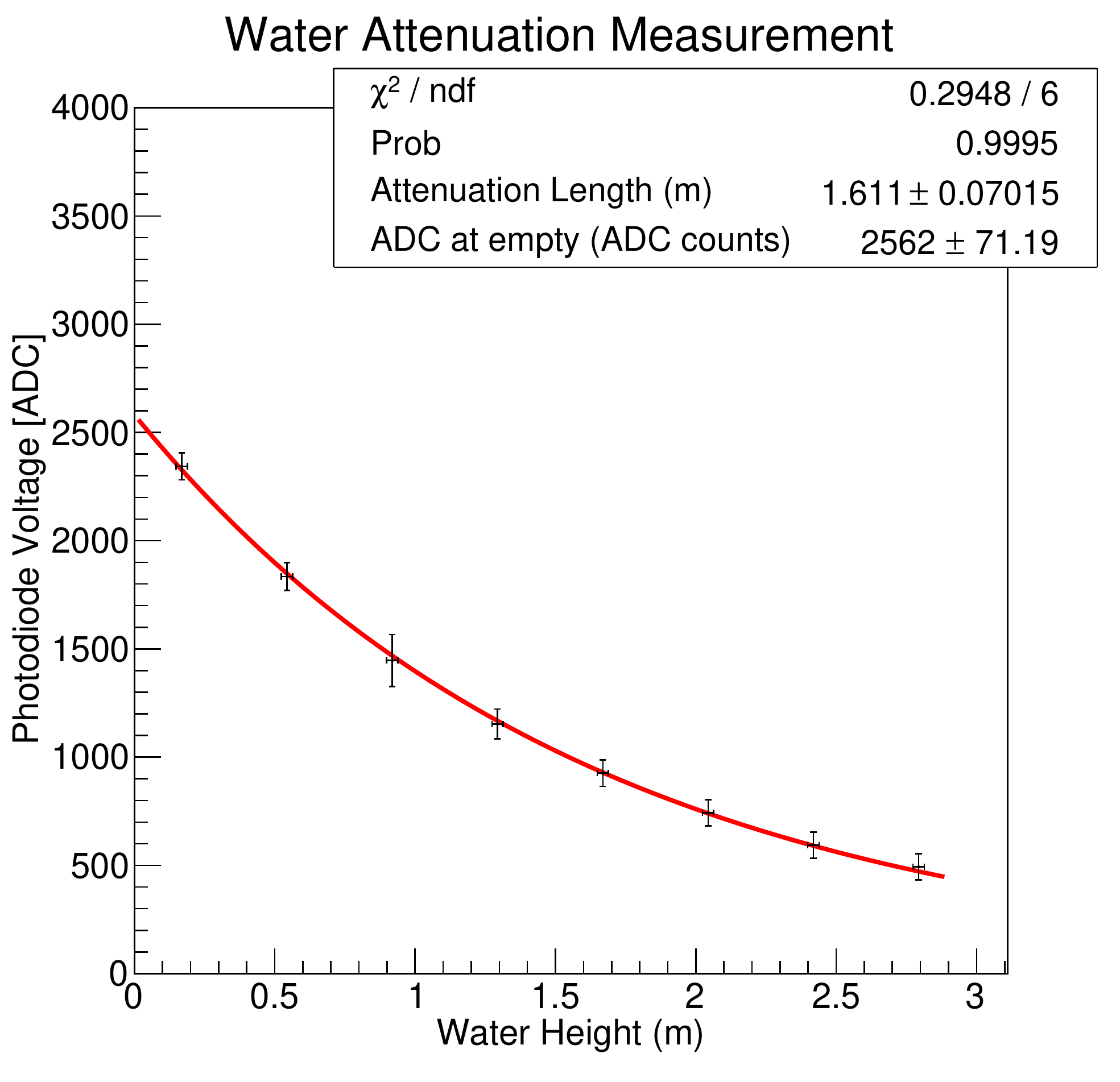}
    \includegraphics[width=0.45\textwidth]{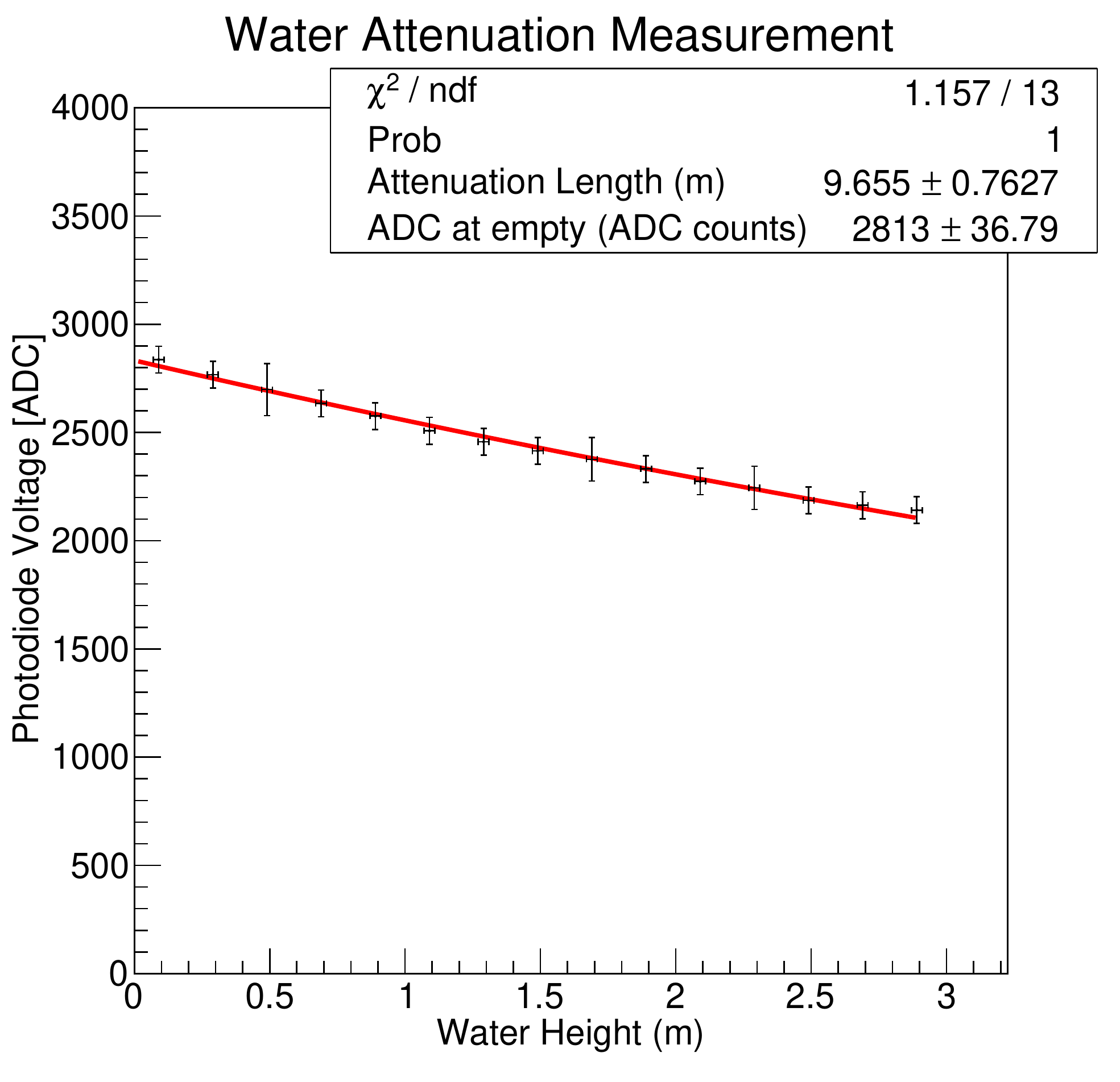}
        \includegraphics[width=0.45\textwidth]{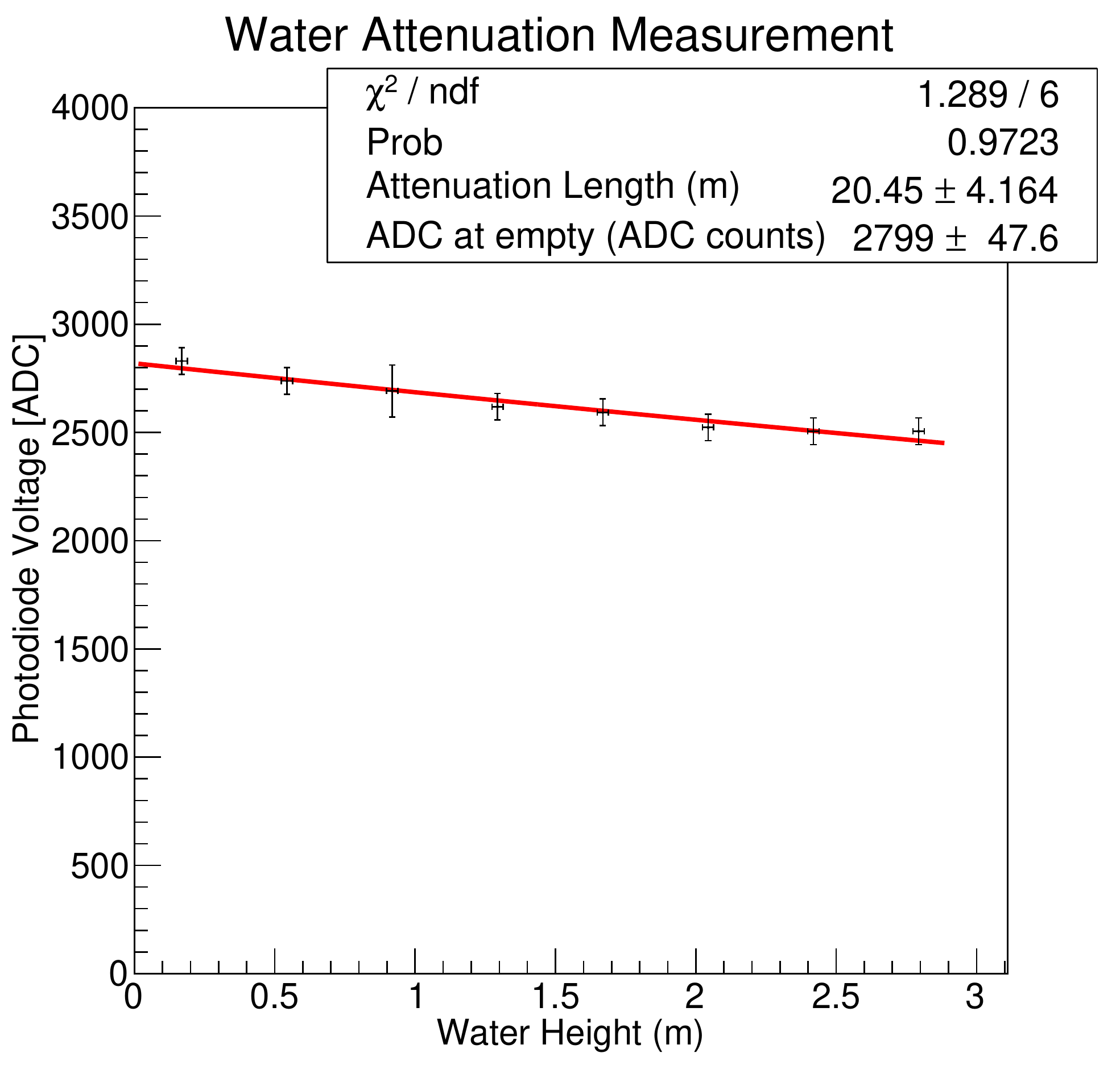}
  \caption{Plot showing water attenuation measurement for raw pit water (top left); after filtering for 12 days (top right) and after filtering for 25 days (bottom)}
  \label{fig:pit_water}
\end{figure}

Once the water was being passed through the filtration system, samples were taken regularly from the barrel to observe how the attenuation length increased with time. The temperature was continuously monitored throughout this procedure and used to estimate the errors on the data points for each attenuation length measurement. 
Initial measurements of the unfiltered pit water can be seen in Figure~\ref{fig:pit_water} (top left), where the attenuation length was measured to be $1.6\pm0.05$\,m. 
The result is in reasonable agreement with the measurement with the Secchi disk for white light. Other results can be seen after 12 days (top right) and after 25 days(bottom). 

\begin{figure}
  \centering
    \includegraphics[width=1\textwidth]{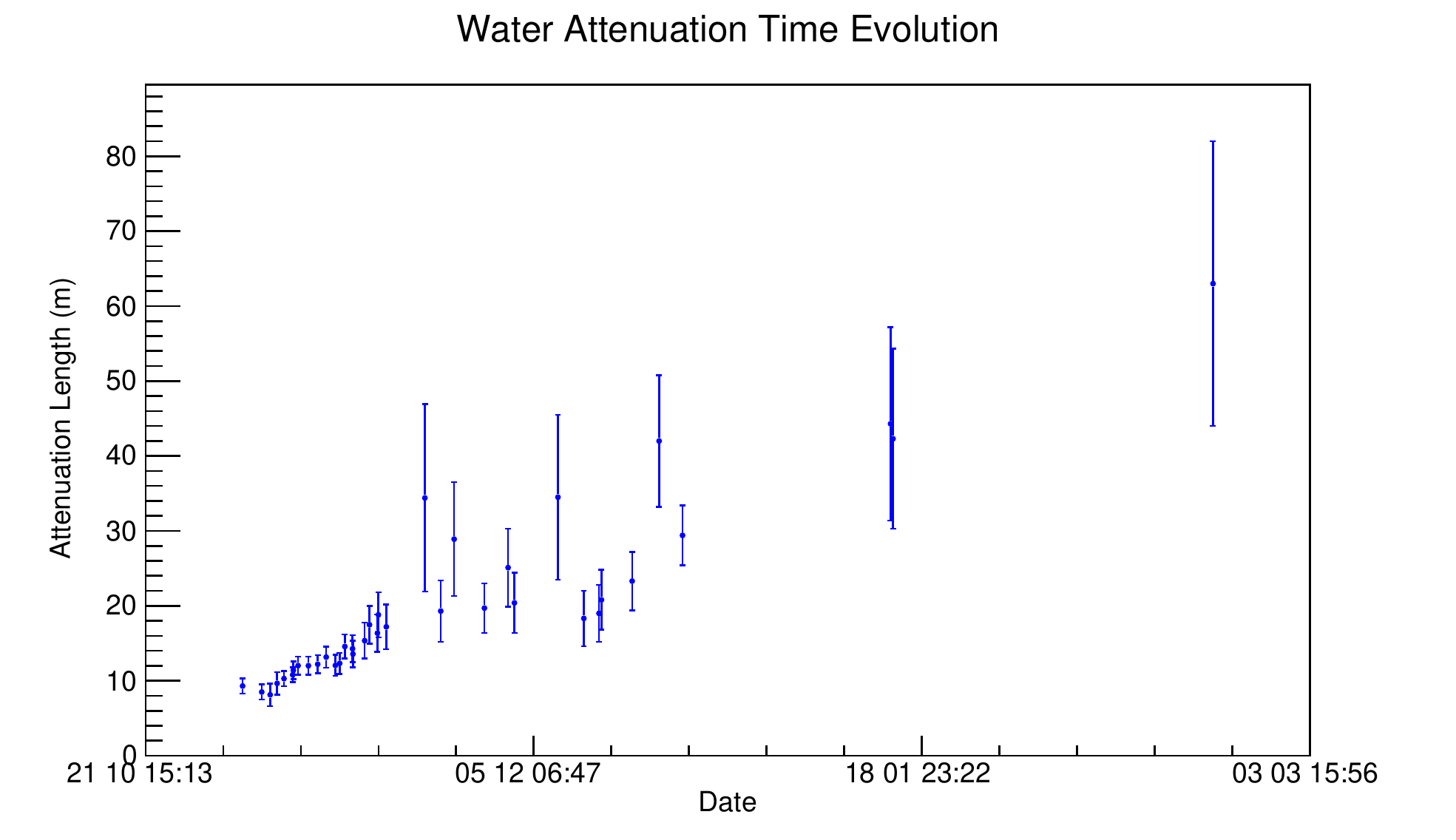}
  \caption{Plot showing how attenuation length improved with time while the water was continuously circulated through the filtration system.}
  \label{fig:time_plot}
\end{figure}

Figure \ref{fig:time_plot} shows the outcome of the measurements over a four month period. Larger errors are associated with measurements during the colder months when the laboratory heater cycles produced large temperature related variations in the ADC output.  The temperature was the dominant systematic delivering an uncertainty over the course of each attenuation length measurement of sometimes up to 38 ADC counts per point for a change between single point measurements of 1$^{\rm{o}}$C. However, it can be seen that the filtration system is effective with a clearly increasing trend in attenuation length up to values of 50-60\,m, and referring back to Figure \ref{fig:illustrate} within the accuracy of the measurement system. This suggests a relatively  simple filtration system will be adequate for a detector size of 30\,m diameter. It can also be seen that a circulation time of about 3-4 months should provide an attenuation length of 50\,m.

\subsection{System Improvements}
\label{sec:improve}
Once the time constant of the water cleaning had been determined, steps were taken to make the setup more stable to improve the accuracy of the measurements at longer attenuation length and to ascertain whether further cleaning would be possible. This was done by insulating the electronics and laser from temperature changes, increasing the water recirculation speed to clean the water much faster, and working in April when the temperature was warm enough to avoid the heater cycles. It is possible that an additional speed up in the water clarity with recycle rate was due to there being no chance for any bacteria to survive such short circulation times, although this is conjecture.

\begin{figure}[h]
\centering
\includegraphics[width=0.9\textwidth]{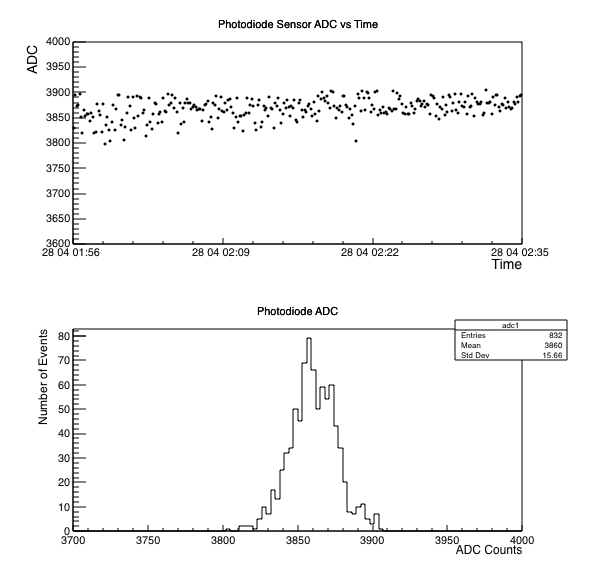}
\caption{Top:  ADC + photodiode stability after system improvements. Bottom: the projection of the ADC values showing an RMS of 15.6 ADC counts.}
\label{fig:ADCstability}
\end{figure}

Figure~\ref{fig:ADCstability} (top) shows the photodiode ADC for a full water column in the improved set up as a function of time. The lower histogram shows the projection with a standard deviation of 15.7 ADC counts. This measurement incorporates all of the systematic uncertainties and results in an overall uncertainty of 0.5\%. While this is smaller than the specification for the laser stability, the measured systematic error of 16 ADC counts is nevertheless applied to each data point in the subsequent attenuation length fit procedure. This is well below the target of 30 ADC counts described above for attenuation lengths above 50\,m.

\subsection{Final Results}
\begin{figure}[t]
  \centering
    \includegraphics[width=0.68\textwidth]{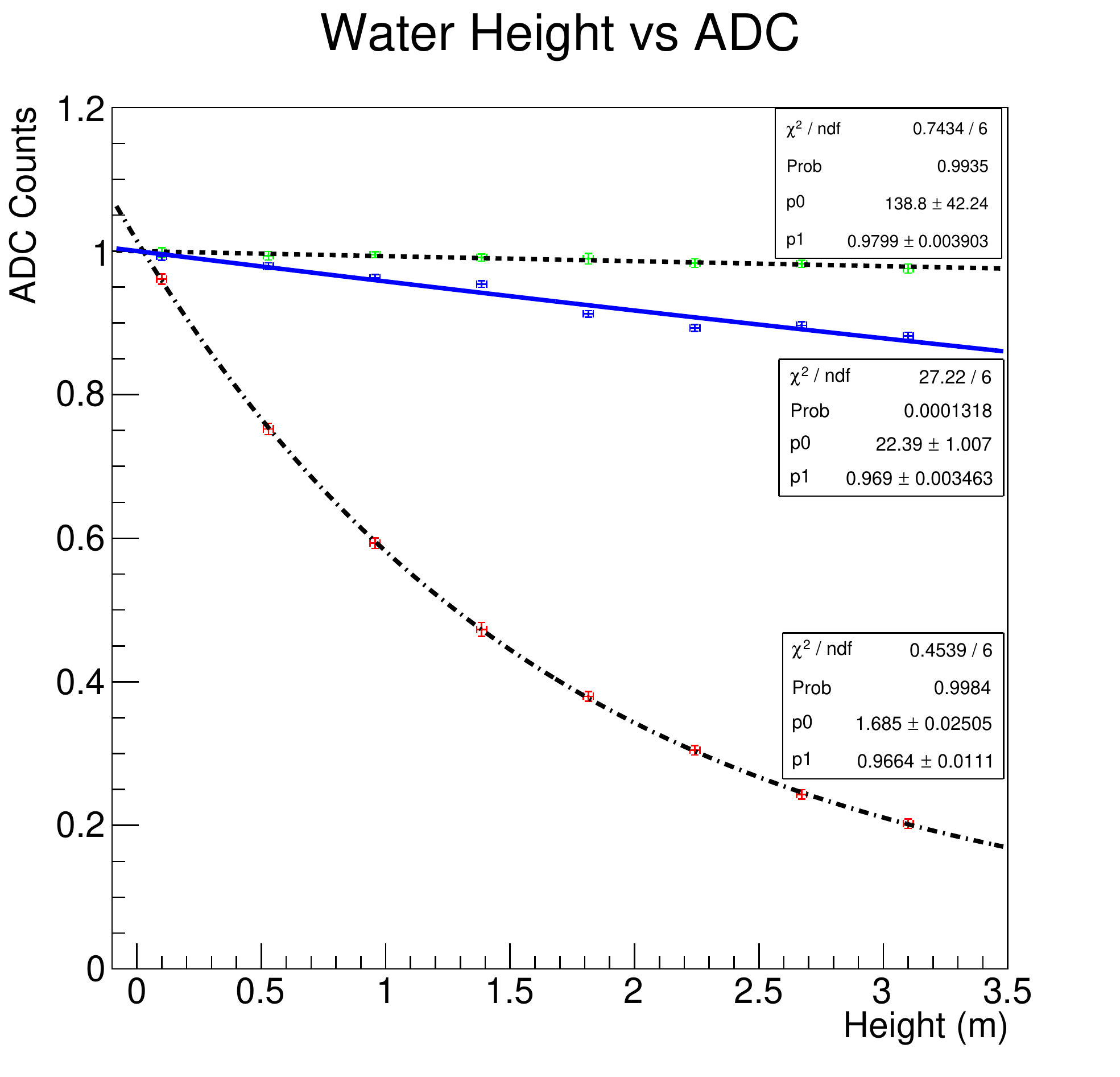}
  \caption{Plot showing attenuation length measurements for three different levels of water clarity, at the start (dot-dash), after 1.5 hours(solid) and after 12 hours(dashed). The y axes have been normalised for comparison to take out longer term drift as described.}
  \label{fig:threeatten}
\end{figure}

Figure~\ref{fig:threeatten} shows three attenuation length measurements taken at intervals of 0, 1.5 and 12 hours of water circulation with 
attenuation lengths of $\lambda$=1.68$\pm$0.02\,m, 22.4$\pm$1.2\,m and 139$\pm$42\,m respectively. This increase in speed of
water cleaning is just due to the increased circulation rate of the water through the filters and sterilizer. 

\section{Simulation}

A dedicated simulation program was developed to cross check and
disentangle any contributions to the attenuation length measurement from the optical setup. 
The simulated photon's trajectory was traced according to the 
rules of geometrical optics and could be absorbed in the water with a probability depending
on the water absorption length. A photon was considered as detected if
it reached the photodetector. The simulation made use of the
ROOT~\cite{root} software framework but all the code describing the photon transportation and
interaction processes was developed independently. 
All the optical elements, including the focussing lens in front of the
photodiode, and the material boundaries present in the
experimental setup were considered in the photon ray-tracing; the laser
beam width and divergence were modelled according to the Thorlabs data sheet~\cite{ref:thorlabs}. The geometrical
parameters  for the laser (direction, starting position)  and the
photodiode (position, size) could all be individually varied.
The water level and its attenuation length could also be varied.
The beam divergence is very small, and measurements made with the above setup were
unable to detect any light more than 5mm away from the laser center. Therefore it was deemed
unnecessary to simulate any scattered light from the inside of the tube as our measurements
showed it would not be possible to detect such a low light level.

Given a random starting position and direction each photon was propagated
through every material boundary, where Snell's Law was
applied, up to the photodetector plane.  The focussing lens was
modelled by applying refraction at the spherical surfaces. 

\begin{figure}[t]
\centering
\includegraphics[width=0.45\textwidth]{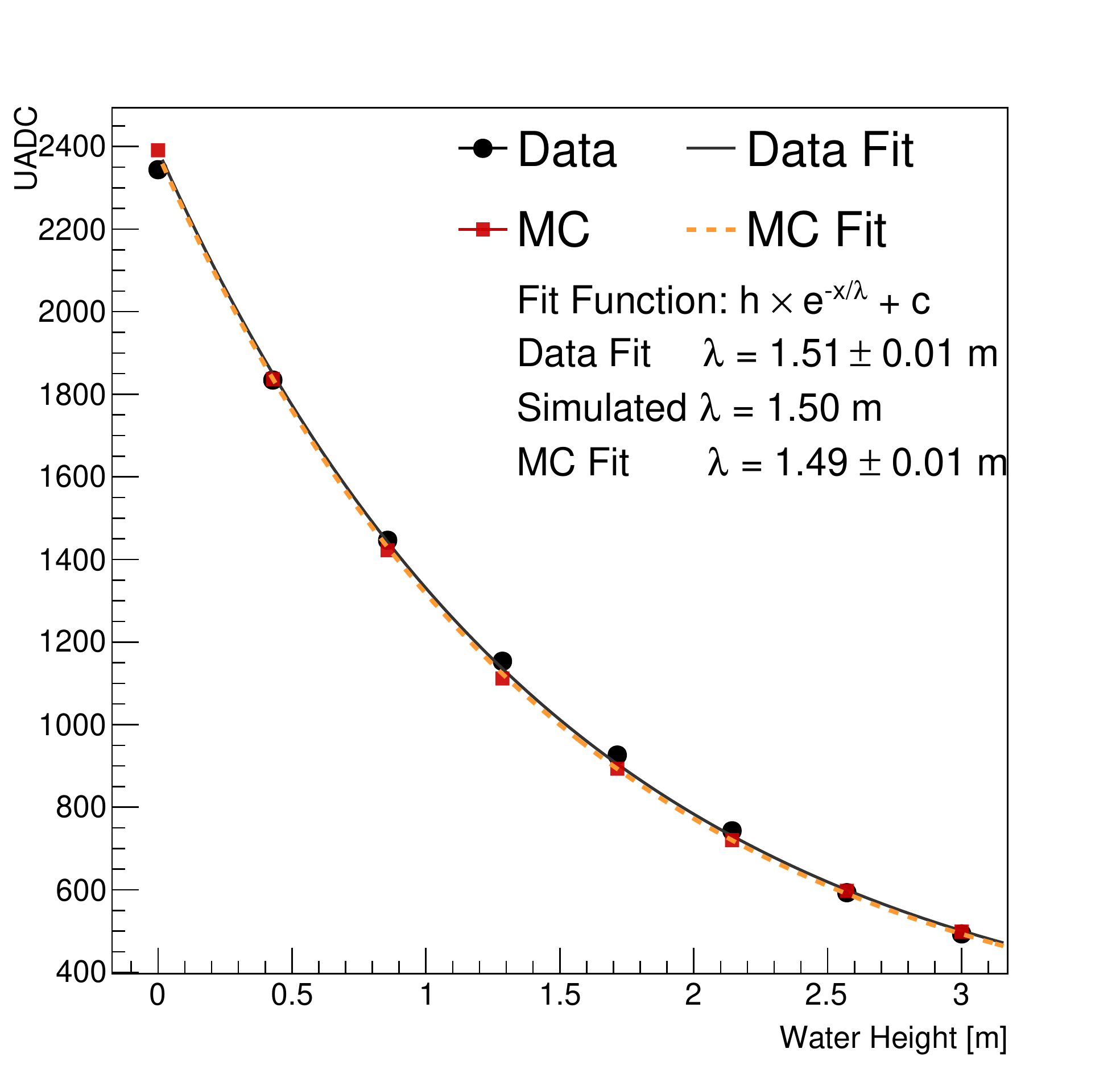}  
\includegraphics[width=0.45\textwidth]{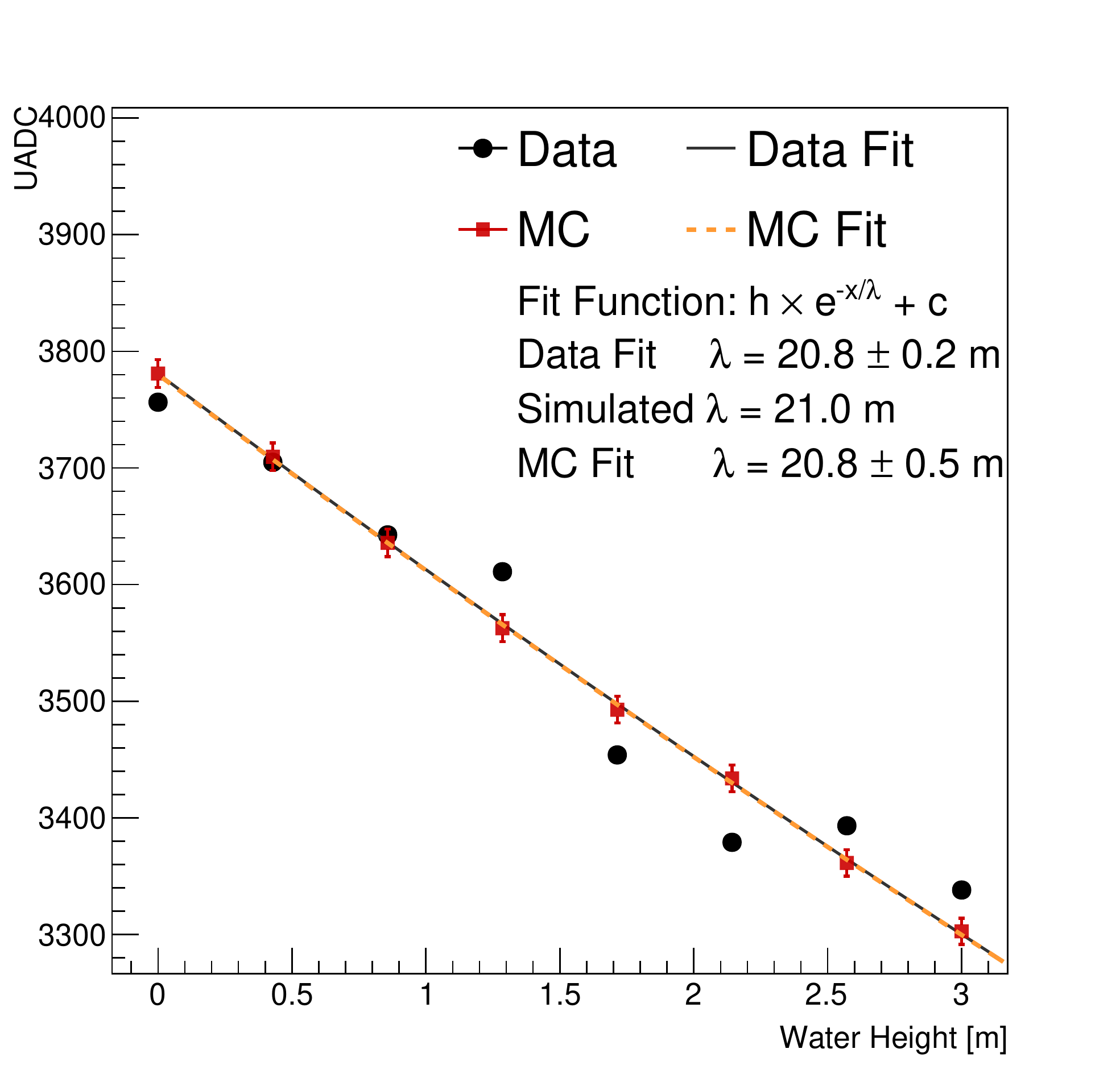}  
\caption{Left: light  intensity  measured by the photodiode expressed in ADC counts, as a
  function of the water level for experimental data (dots) and simulation
  (squares). The attenuation length ($\lambda$) extracted from a fit (solid line) 
  to the data  points is $\lambda =  1.51 \pm 0.01$\,m. Also shown is the 
  simulation with an input value of  $\lambda =  1.5 $\,m whose corresponding
  fit (dashed line) gives $\lambda =  1.49 \pm 0.01$\,m. Statistical errors only are used.
Right: a longer attenuation length of  $\lambda =  20.8 \pm 0.2 \,$m for the data and the input to the simulation 
of  $\lambda =  21 \,$m shows the corresponding fit of $\lambda =  20.8 \pm 0.5$\,m.}
\label{fig:simfigs}
\end{figure} 
The simulated data points and the fit to them showed there were no unexpected systematics associated with
the optical geometry: for example the refractive index changes from air to perspex, to water, to air, to glass and to lens
did not produce any effect which could be mistaken for a change in attenuation length. The simulation results are
summarised in Figure \ref{fig:simfigs} where it is shown that the
relative attenuation length fits are both in good agreement with the results from the experimental data and demonstrate 
no unaccounted for systematic effects for both short and medium attenuation length.

\section{Conclusions}
The attenuation length of the water from the Wentworth Pit 2W, site of the CHIPS experiment, has been measured
during filtration. The UV sterilization and simple filtering method was found to be capable of delivering in the region of 100\,m
attention length at a wavelength of 405\,nm. 
This is completely adequate for a 30\,m diameter CHIPS detector as shown
by detailed simulation and reconstruction studies. Without the need for the exclusion of dissolved solids,
the operational complexity and cost of the water treatment system is minimised. 
Furthermore, the density difference between inside and outside the detector is small leading to reduced requirements on the
mechanical strength of the lightweight structure and associated cost implications.

\section{Acknowledgements}
This work has been carried out with the support of the DOE (US), The Leverhulme Trust (UK) and A*MIDEX project (n¡ ANR-11-IDEX-0001-02) funded by the ÇInvestissements dÕAvenirÈ French Government program, managed by the French National Research Agency (ANR), and the US NSF. Especial thanks go to the mine crew of the Soudan Underground Laboratory. The authors would like to thank M. Smy and H. Sobel at the University of California at Irvine for constructive discussions.

%\end{linenumbers}
\bibliographystyle{elsarticle-num}
\bibliography{waterbib}

%\begin{thebibliography}{99}
%\bibitem{ref:fnalloi}Cherenkov detectors In mine PitS (CHIPS) Letter of Intent to FNAL, arXiv 1307.5918 (2013)
%\bibitem{ref:secchi}https://en.wikipedia.org/wiki/Secchi\_disk
%\bibitem{ref:leighandandy}Neutrino 2016 proceedings, J.Phys.: Conf. Ser. to be published.
%\bibitem{ref:waterturnover} Clay's Handbook of Environmental Health, p272
%\bibitem{ref:michealsmy}Micheal Smy, priviate communication 2015.
%\bibitem{ref:BBB}https://beagleboard.org/black
%\bibitem{root} R. Brun, F. Rademakers,{``ROOT - An Object Oriented Data Analysis Framework''},
%Proceedings AIHENP'96 Workshop, Lausanne, Sep. 1996, Nucl. Inst.\& Meth. in Phys. Res. A 389 (1997) 81-86. 
%See also http://root.cern.ch/.
%\bibitem{ref:thorlabs} https://www.thorlabs.com/thorproduct.cfm?partnumber=CPS405
%\end{thebibliography}
\end{document}